\documentclass[journal]{IEEEtran}
\ifCLASSINFOpdf
\else
\fi
\usepackage{graphicx}
\graphicspath{ {images/} }

\usepackage{enumitem}

\usepackage{verbatim}
\usepackage{comment}

\usepackage{tabu}
\usepackage{booktabs}
\usepackage{caption}
\usepackage{subcaption}

\usepackage[hidelinks]{hyperref}
\usepackage{xcolor}

\newcommand{ \textinserted } [1]{ #1} 
\newcommand{ \textmodified} [2]{ #2} 
 \newcommand{ \textremoved} [1]{ } 

\begin{document}
%
% paper title
% Titles are generally capitalized except for words such as a, an, and, as,
% at, but, by, for, in, nor, of, on, or, the, to and up, which are usually
% not capitalized unless they are the first or last word of the title.
% Linebreaks \\ can be used within to get better formatting as desired.
% Do not put math or special symbols in the title.
\title{Polarity in the Classroom: A Case Study Leveraging Peer Sentiment Toward Scalable Assessment}
%
%
% author names and IEEE memberships
% note positions of commas and nonbreaking spaces ( ~ ) LaTeX will not break
% a structure at a ~ so this keeps an author's name from being broken across
% two lines.
% use \thanks{} to gain access to the first footnote area
% a separate \thanks must be used for each paragraph as LaTeX2e's \thanks
% was not built to handle multiple paragraphs
%
%
%\IEEEcompsocitemizethanks is a special \thanks that produces the bulleted
% lists the Computer Society journals use for "first footnote" author
% affiliations. Use \IEEEcompsocthanksitem which works much like \item
% for each affiliation group. When not in compsoc mode,
% \IEEEcompsocitemizethanks becomes like \thanks and
% \IEEEcompsocthanksitem becomes a line break with idention. This
% facilitates dual compilation, although admittedly the differences in the
% desired content of \author between the different types of papers makes a
% one-size-fits-all approach a daunting prospect. For instance, compsoc 
% journal papers have the author affiliations above the "Manuscript
% received ..."  text while in non-compsoc journals this is reversed. Sigh.

\author{Zachariah J. Beasley, Les A. Piegl, and Paul Rosen}% <-this % stops a space
\maketitle

% As a general rule, do not put math, special symbols or citations
% in the abstract or keywords.
\begin{abstract}
Accurately grading open-ended assignments in large or massive open online courses (MOOCs) is non-trivial. Peer review is a promising solution but can be unreliable due to few reviewers and an unevaluated review form. To date, no work has 1) leveraged sentiment analysis in the peer-review process to inform or validate grades or 2) utilized aspect extraction to craft a review form from what students actually communicated. Our work utilizes, rather than discards, student data from review form comments to deliver better \textit{information} to the instructor. In this work, we detail the process by which we create our domain-dependent lexicon and aspect-informed review form as well as our entire sentiment analysis algorithm which provides a fine-grained sentiment score from text alone. We end by analyzing validity and discussing conclusions from our corpus of over 6800 peer reviews from nine courses to understand the viability of sentiment in the classroom for\textmodified{increasing the reliability and scalability of open-ended assignments in large courses}{increasing the information from and reliability of grading open-ended assignments in large courses.}
\end{abstract}

% Note that keywords are not normally used for peerreview papers.

% For peer review papers, you can put extra information on the cover
% page as needed:
% \ifCLASSOPTIONpeerreview
% \begin{center} \bfseries EDICS Category: 3-BBND \end{center}
% \fi
%
% For peerreview papers, this IEEEtran command inserts a page break and
% creates the second title. It will be ignored for other modes.
\IEEEpeerreviewmaketitle

\section{Introduction}
%Our project began by discovering something very obvious from students: they were bored. Traditional lectures were no longer appealing to students. In an effort to not hold back students, we decided to give them more control over their education, both on the teaching and critiquing end. This work deals with the later.\\

Scalable grading of complex assignments (e.g., websites, essays, designs, open-ended questions) is notoriously hard in large, online academic environments \cite{shah2014some}--\cite{beasley2018ten}. Peer review is an often-cited solution, but can be unreliable due to a low number of reviewers and an unevaluated review form. Our work addresses both issues in a novel way by leveraging sentiment analysis in the peer-review process. By organizing our courses to include peer teaching and crowdsourced peer review (as in \cite{beasley2019designing}), we obtain a crowd of reviewers large enough (30--40 on average, although we have found 20 to be sufficiently stable) to confidently assess a student's work and as a side effect stimulate learning in reviewing students \cite{russell2017variability}--\cite{li2019does}.
% 2-3 \cite{singh2016question}, \cite{kulkarni2014scaling}
% 7 \cite{almatrafi2018systematic}
% three references - twice

\begin{figure*}[h]
\centering
\includegraphics[scale=0.50]{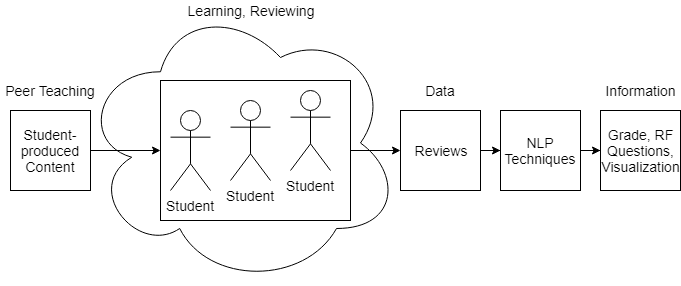}
\caption{Assessment process highlighting when natural language processing techniques are utilized.}
\label{fig:assess}
\end{figure*}

Rather than apply natural language processing techniques to a work itself (e.g., an automated essay scorer like \cite{somasundaran2018towards}), we apply the techniques to the meaningful content generated by a peer-reviewing crowd (\autoref{fig:assess}). Our lexicon-based approach allows us to 1) collect sentiment to derive a fine-grained grade of peer-review text, 2) verify that grade with various metrics, and 3) suggest aspects important to the reviewers: candidates for addition to the review form.\textinserted{In this way, we can fulfill a ``need for data collection tools that support teaching and learning'' in engineering courses \cite{moyne2018development}.} Our data-driven process addresses this need by providing the instructor with a better understanding of peer review and by increasing confidence in assigning a grade for complex assignments in large and massive open online courses (MOOCs).
 %Our data-driven process allows us to gather the maximum amount of information from the peer reviewer process, which can provide the instructor with a better understanding of a peer review,
 
The main contribution of this work is the description of a novel \textit{process}: the development of a domain-specific lexicon and sentiment scorer to equip an instructor to build a review form through aspect extraction and to grade peer-review comments through sentiment analysis. It is not a one-size-fits-all review form or lexicon, but should be tailored to a specific course context.\textinserted{The lexicon-based approach was chosen primarily to improve the information with which numeric results can be justified.}

%Finally, for the reviewing student, the process has the added benefit of promoting learning \cite{russell2017variability}, \cite{li2019does}.

%
%As we opened up our courses to peer review, we discovered that we were not sure what students were looking for in a quality submission. So we began a data-driven process to find out what was important. We started with a rough draft of a review form, opened it up to detailed responses, and gradually honed in on a student-produced review form after collecting students' comments over multiple semesters. Although an instructor-produced review form may tell the instructor what he/she is looking for, it may not communicate what a student's peers think --- in our belief, the primary purpose of peer review. Thus, while we continue to mine for new review form questions, we also leverage sentiment from student detailed comments. This allows us to gather the maximum amount of information from peer reviewers, which can provide the instructor with a better understanding of a peer review, increase confidence in assigning a grade, and ultimately move towards removing the instructor/TAs as the grading bottleneck of open-ended assignments in massively open online courses (MOOCs).

\section{Related Work}
Because sentiment analysis is widely used to predict ratings from text in a large variety of domains, we attempt to briefly cover representative examples from social media, creative works (e.g., movies), and products. In an ideal world, we could simply take the best trained classifier from one of these categories and use it to determine student grades. However, there are a number of complications to this approach detailed below. When highlighting others' approaches, we report on either accuracy, precision, recall, or F1 score according to the evaluation prioritized in the author's own work.

%Accuracy refers to the percentage of examples classified correctly. Precision is the number of true positives over the sum of true positives and false positives (i.e., the ability to prevent false positives). Recall is the number of true positives over the sum of true positives and false negatives (i.e., the ability to find all true positives). The F1 score is the harmonic mean of (giving equal importance to) precision and recall: \[2*\frac{precision * recall}{precision + recall}\]

\subsection{Transfer Learning} \label{sec:transfer}
%For example, taking the best sentiment classifier in movie reviews and utilizing it in the educational domain. 
Transfer learning in the field of machine learning is the process of applying an algorithm in a different domain than that in which it was trained. From an ethical standpoint, basing student grades on a classifier from a different domain could result in unpredictable and unfair results. From a practical standpoint, sentiment analysis is ``highly sensitive to the domain from which the training data is extracted" \cite[p. 31]{liu2012sentiment}. Others have found wide disagreement in results when applying the same publicly available sentiment analysis tools to different domains (e.g., accuracy of 80.4\%, 62.5\%, and 71.5\% on movie reviews, tweets, and Amazon reviews, respectively) \cite{serrano2015sentiment}.

Still, various approaches to domain adaptation have been attempted. Most require labeled data from at least the source domain (e.g., \cite{blitzer2007biographies} and \cite{pan2010cross} which link based on related features and related words, respectively). Others address this task when there is limited labeled data and note performance improvements over ignoring unlabeled data, but still do not address a situation such as ours with no labeled data \cite{goldberg2006seeing}. Another alternative is a nearly unsupervised approach based on a two-phase clustering method, but a domain-dependent lexicon is still needed or else accuracy will ``suffer a serious performance loss once the domain boundary is crossed'' \cite[p. 269]{mudinas2018bootstrap}. A hybrid lexicon with weighted values from a local and general-purpose lexicon was utilized by \cite{muhammad2016contextual}, but performance improvements were found on only two out of three social media datasets (with a much lower F1 score on the third). In fact, over half of the combinations they tried decreased the F1 score, even though all lexicons were in the social media domain. Finally, there has been observed a ``clear benefit to creating hand-ranked, fine-grained, multiple-part-of-speech dictionaries for lexicon-based sentiment analysis" \cite[p. 282]{taboada2011lexicon}.

In their paper testing 24 algorithms on 18 datasets (including Twitter, Yelp, YouTube, Amazon, Digg, BBC, TED, and Myspace) to determine benchmarks in sentiment analysis, \cite[p. 23, 27]{ribeiro2016sentibench} note that ``sentiment analysis methods cannot be used as `off-the-shelf' methods, especially for novel datasets'' and found that ``methods are often better in the datasets [in which] they were originally evaluated'' even for popular algorithms like SentiStrength \cite{thelwall2012sentiment}, \cite{thelwall2010sentiment} and SO-CAL \cite{taboada2011lexicon}. Others have found significant improvement with a crowd-labeled, domain-specific negative word lexicon compared to two other general-purpose lexicons classifying negativity in German media reports and party statements \cite{haselmayer2017sentiment}. They, like us, strongly advocate a \textit{process} and note that even ``some commercial providers advise against using their sentiment lexicon out-of-the box without customizing it to the domain'' \cite{haselmayer2017sentiment}. Specifically, when narrowed down to sentiment analysis algorithms in the software engineering domain (Stack Overflow, Jira, etc.), \cite{novielli2018benchmark} claim that publicly available sentiment analysis tools are inadequate and disagree with one another, but that domain-specific tools enhance accuracy. Others even go so far as to label the current state-of-the-art sentiment analysis tools deficient on the software engineering domain \cite{lin2018sentiment}. Taken together, this research shows the necessity of a domain-dependent lexicon and algorithm for accuracy in educational sentiment analysis.

\subsection{Sentiment Analysis in Education}
Influential and widely cited works summarizing sentiment analysis tools and techniques have often lacked any reference to its applications in academia \cite{liu2012sentiment}, \cite{feldman2013techniques}, \cite{cambria2013new}.\textmodified{However, in the past three to four years papers have begun to emerge leveraging sentiment analysis in academia, although to date it has not been applied to generate a student grade based on review comment text.}{However, in the past three to four years papers have begun to emerge leveraging sentiment analysis in academia, although to date it has not been applied to generate a grade based on review comment text or to provide any additional information to the \textit{instructor}.} Similarly, we find no research leveraging aspect extraction to semiautomate review form creation in a data-driven way.\textinserted{The closest conceptual work to developing a review form from student input or data is a small-scale study in which 25 nursing students created a ``a learner-driven feedback form" to reduce ``the possibility of negative peer review experience" \cite[p. 47]{duers2017learner}.}\textmodified{In any case, it is important to gather the maximum information from educational data, as businesses, artists, and politicians have discovered in other domains.}{In any case, it is important to improve both the quantity and quality of the information gathered from assessment data to allow instructors to pursue data-driven decision-making in the classroom.}

Sentiment analysis has recently been used in the classroom to determine:
\begin{itemize}
\item student attrition over time in three MOOCs (captured from course forums and scored by a product reviews lexicon \cite{wen2014sentiment}) or predicted attrition in a single MOOC (captured from a course forum and scored by SentiWordNet 3.0 as one feature of a neural network \cite{chaplot2015predicting});
\item the mood of students toward a teacher (captured via Twitter and scored by Naïve Bayes \cite{esparza2016proposal}) or as an ``emotional thermometer for teaching'' in virtual classrooms (captured in forum posts and scored by an ensemble \cite{pousada2017towards});
\item negative students or course issues (captured in online course forums and scored by the Microsoft Text Analytics API \cite{schubert2018measuring} or captured from social media and scored by a mixed graph of terms \cite{clarizia2018learning});
\item teacher strengths and weaknesses identified by students (a proposed system with sentiment captured via questionnaire and scored by Naïve Bayes \cite{balahadia2016teacher}, or a proposed multilingual system with sentiment captured from Coursera peer reviews and scored by a lexicon in R \cite{rani2017sentiment}, or a system with sentiment captured from teacher evaluations and scored by an ensemble \cite{lalata2019sentiment});
\item student perception of internship experience (captured from transcribed interviews and scored manually \cite{fleming2018exploring});
\item an alternative way to view poetry and a means of student discussion on the relationship between text and numbers (captured from a Walt Whitman poem and scored by a proprietary sentiment analysis tool \cite{lynch2015soft}).
\end{itemize}

\textinserted{There is one recent work which applies sentiment analysis to peer-review text, but it is limited to excluding highly negative reviews from being provided to the submitters so they will not disregard and disengage from the formative feedback they receive \cite{gehringer2019board}. It uses the VADER lexicon \cite{gilbert2014vader} to score content, which we have found inaccurate on our dataset when compared with other lexicons \cite{beasley2021domain}. In addition, this work provides no other metrics or visualizations to the instructor and it does not contribute to a scalable, reliable grading process.}\textremoved{They do not utilize information from students to develop a review form. }\textinserted{While these applications of sentiment analysis in the classroom are interesting and helpful in their own ways, they do not contribute to an information gain from peer review toward improving the review form or providing a reliable grade.} None go further than classical applications of sentiment analysis, merely noting subjective opinion toward an object.

\section{Methodology}
%Our review form was comprised of two components: an analytical section with questions and radio button responses and a subjective, free response section where the student was encouraged to comment on any aspect of the work being reviewed. Both components were weighted and contributed to a student's final grade.

\textmodified{Our peer-review data (individual review form responses both analytical [radio button and subjective) came primarily from software engineering (SE), software testing (ST), computer graphics (CG), and geometric modeling (GM) courses over a period of five semesters.}{Our peer-review data was comprised of individual review form responses containing both analytical (radio button) and subjective (essay question) feedback. It came primarily from software engineering (SE), software testing (ST), computer graphics (CG), and geometric modeling (GM) courses at a large, R1 university computer science department over a period of five semesters.} Of the nine courses, four were undergraduate only, the rest were cross listed for graduate and undergraduate students.\textinserted{Each course had approximately 40 students.}

%, and data visualization (DV)

From these courses, we have a number of growing corpora of over 6800 peer reviews of 325 student works. The responses reflected peer's sentiment on three different types of projects:\textmodified{a (weekly, 35--40 reviewers) group presentation, a (semester, 20--40 reviewers) group essay, and a (semester, 20--40 reviewers) group term project.}{a weekly group presentation (35--40 reviewers), a semester-long group essay (20--40 reviewers), and a semester-long group term project (20--40 reviewers).}\textinserted{Compared to other peer-review analyses, the number of reviews per student work was quite large \cite{gehringer2019board}, \cite{piech2013tuned}.} Textual responses were aggregated by our system to provide the mean, median, and standard deviation of sentiment, number of comments successfully scored, and various per-comment metrics (Section \ref{grade}).
%), 2) over 2,200 peer reviews from three DV courses (840 student works), and 3) over 2,500 peer reviews from \href{https://www.peerlogic.org/}{PeerLogic (PL)}, an open repository of several peer review systems. These corpora give us a variety of perspectives on the peer review process, since each instructor utilizes peer review in different ways.

%In the SE, ST, CG, and GM courses,  In the DV courses, the peer reviews were performed roughly bi-weekly on a semester-long individual visualization programming project by 2-4 reviewers. The PL peer reviews contained no information on the type of project or number of reviewers. 

\subsection{Review Form: Analytical Feedback}
The analytical section of our review form was created using an iterative, data-driven approach
 \cite{beasley2019designing}. 
Our seed growing algorithm began with a basic rubric and questions were added, modified, or removed after each semester through intelligent data combing: \textit{a process of selecting information-rich keywords and phrases, through human intelligence, for the purpose of correctly analyzing and summarizing student observations}. This process was intentionally fuzzy---words were selected if they provided 1) meaningful sentiment (e.g., ``extraordinary,'' but not ``good'') or 2) information (e.g., students mentioned a presence or lack of ``citations'' or ``diagrams''). This was later semiautomated through the use of an aspect extractor detailed in Section \ref{sec:aspect_extractor}, but still requires human intelligence for verification. The current review form has 22 questions divided into three categories: Overall score, Technical score, and Personalization score. Each question captures a student's response in an area that previous students have indicated is important \cite{piegl2019assessing}.

\subsection{Sentiment: Subjective Feedback}
%Even though we mined questions and answers from students' detailed comments, we may not have reached an exhaustive list. Thus, 
Selecting a review form radio button communicates a little information but allowing detailed feedback provides another dimension of student response: sentiment. Though the analytic portion of the review form restricted students, the subjective section allowed freedom to discuss anything. Sentiment, along with a concise summary of what students actually said, provided rich information for an instructor to validate a peer-review score, particularly if the course was large enough that the instructor could not check every work. Thus, the sentiment score aggregated over the crowd's response was a component of our final score that could be used to validate or adjust the grade from the analytic section.

%Furthermore, just as communication is more than the words we speak (body language, tone, etc.), merely clicking a radio button communicates little information. Allowing detailed feedback provides another dimension of student response: sentiment. Though the analytic portion of the review form restricted students, the subjective section allowed freedom to discuss anything. Sentiment, along with a summary or visualization of what students actually said, provided rich information for an instructor to validate a peer score (particularly if the course is large enough that the instructor cannot check every work). Thus, this sentiment score aggregated over the crowd's response was a component of our final score that could be used to validate or adjust the grade from the analytic section.

\subsection{Assumptions} \label{assume}
In our process, we aggregated document-level sentiment using an opinion lexicon (semantic orientation is according to instructor heuristic) for our domain. We had a regression problem on unlabeled data and presented an ensemble of scores to the instructor for reference, utilizing a soft voting scheme. We reported on regular (not comparative), direct (not indirect), and explicit (not implicit) opinions \cite{liu2012sentiment}.

The final major assumption in our work was that we could not utilize an instructor grade as a ground truth for \textit{every} individual's textual response. Many student reviewers, learning the material for the first time, presumably commented on fewer aspects of the work than the instructor, who had a better sense of the big picture. It was also possible that a student found something the instructor missed. Ultimately, we could not expect each student's review to conform to the instructor's. This prevented us from utilizing any grading system that measured reviewer competency (e.g., \cite{piech2013tuned}), trained reviewers to review like an instructor, or penalized ``poor" reviewers. The aggregate peer-review score was meant to accurately reflect the opinion of the crowd that consumed the information presented, although it could be adjusted by the instructor.\textmodified{Not using the instructor score as a ground truth safeguarded against a biased system, but required thinking outside the box to measure reliability.}{Not using the instructor score as a ground truth also safeguarded against introducing bias into the algorithm.} Our assumption of unlabled data also motivated the choice of a lexicon-based approach in addition to its intuitiveness, interpretability (since rationale for a grade was occasionally requested), and accuracy on short segments of text (see \cite{belinkov2019analysis} for some other limitations of neural networks, especially identifying which linguistic properties are identified and explaining predictions).\textinserted{Ultimately, rather than try to use an existing recurrent neural network trained on a different labeled dataset (e.g., \cite{yang2019xlnet}), we chose an approach that would increase the amount of information available to the instructor.}

%\subsection{Data}
%To date, we have over 6,800 data points from individual review form responses (analytical and subjective) by peers in the same course. The responses reflected peer's sentiment on three different types of projects: a (weekly, 35-40 reviewers) group presentation, a (semester, 20-40 reviewers) group essay, and a (semester, 20-40 reviewers) group term project. All textual responses were aggregated to provide the mean, median, and standard deviation of sentiment, number of comments successfully scored, and various per-comment metrics (\autoref{grade}).

\section {Sentiment Analysis}

\subsection{Lexicon and Review Form} \label{sec:aspect_extractor}
% We continued to mine questions and answers from students' detailed comments every semester to add to or modify the review form until we reached saturation (i.e., potential new topics were used very infrequently or only added redundancy). 

Over the course of five semesters, we gathered keywords for our lexicons through the process of intelligent data combing described above and weighted them by instructor heuristic \cite{taboada2011lexicon}, \cite{afinn}. In contrast to the review form questions, which were selected for their breadth, we selected any words that exhibited positive or negative sentiment. However, we intentionally excluded overused words like “good” and “bad” that provide little quality information---\cite{pang2008opinion} also found a balance of frequent and rare words necessary to discover subjective content. Thus, our lexicons could be interpreted as a stemmed seed set which was not expanded through a lexical learning strategy since we desired a smaller set of words specific to our domain and weighted by heuristic (in contrast to  \cite{hatzivassiloglou1997predicting}--\cite{hu2004mining}).\textinserted{Lemmatization was considered, but stemming was chosen for its ability to reflect more closely the actual written text. It allowed identification of keywords at a finer level of detail, including superlative adjectives (e.g., a ``dry" vs. the ``driest" lecture, which should be weighted differently). Similarly, the keywords ``good," ``better," and ``best" all have the same lemma: ``good." Thus, ``better" and ``best," which do exhibit significant sentiment, would have been excluded from analysis. Indeed, there would have been an average of 151 such cases per course in Spring 2019 of these two words alone being excluded if we had used lemmatization. Great care was taken when creating the lexicon to consider words in their variety of inflected forms.}
% three references
%, \cite{turney2002thumbs}, and 

Our positive word lexicon currently contains 250 words and our negative word lexicon currently contains 187 words. We have an additional lexicon comprised of words that negate sentiment (19 words) and a lexicon for flag words like ``cheating'' (12 words).\textinserted{The inclusion of flag words allows the potential for crowdsourced plagiarism detection \cite{beasley2020crowd}. When compared both qualitatively and quantitatively to six other lexicons publicly available (AFINN-111, ANEW-2017, MPQA, SentiWordNet 3.0, SlangSD, and Vader), our domain-specific lexicon provided more consistent tagging of high quality sentiment while appropriately ignoring neutral text \cite{beasley2021domain}, \cite{beasley2020helps}.}

%We leave comparing our lexicon to other general-purpose lexicons \textit{within our context} as a later work.

\begin{figure}[h]
\centering
\includegraphics[scale=0.49]{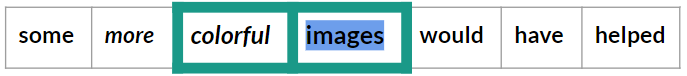}
\caption{Example sliding window over a review comment.}
\label{fig:slidingWindow}
\end{figure}

To semiautomate our process of intelligent data combing, we developed an aspect extractor similar to \cite{blair2008building} that used a sliding window (\autoref{fig:slidingWindow}) around sentiment-laden text (adjectives -- italicized) to suggest aspects (nouns -- highlighted) in close proximity for addition to the review form. This is a form of association rule mining between a noun and a set of sentiment words.\textinserted{There may be multiple---perhaps competing---sentiments on a single aspect. For example, in the review ``...presentations were informative but dry...," both ``informative" and ``dry" are adjectives, with opposite polarity, associated with the noun ``presentations." By finding the target of each sentiment word, rather than starting with the aspect, we were able to fine-tune the total sentiment per aspect.} In \autoref{fig:slidingWindow}, an adjective and noun match is found within the window.\textinserted{At the end of each semester, aspects were analyzed and if an aspect met a customizable threshold of mentions and absolute sentiment, it was considered a candidate for the review form and provided to the instructor (\autoref{table:exampleAspects}). This process was also used to validate the current review form questions, which were all nouns, and to detect ``parroting"---students simply mentioning keywords from the review form itself.}  Although bootstrapping the review form questions and lexicon by hand could not be avoided (and was in fact desirable for accuracy and intelligibility), the aspect extractor assisted with further iterations. Our aspect extractor focused on explicit aspects and ignored implicit aspects (in contrast to \cite{hu2004mining}).\textinserted{We evaluated our aspect extractor with the SemEval Aspect-Based Sentiment Analysis restaurant dataset, surpassing the highest precision and experiencing moderate recall in the new domain \cite{beasley2020sentiment}.}

\begin{comment}
\begin{figure*}[h]
\centering
\includegraphics[scale=0.44]{ae_example}
\caption{Example aspects from an entire semester.}
\label{fig:aeExample}
\end{figure*}
\end{comment}

\begin{table*}[!t]
\centering
\caption{\\Example Aspects from an Entire Semester}
\label{table:exampleAspects}
\resizebox{2\columnwidth}{!}{%
   \begin{tabular}{@{} l*{3}{l} @{}}
      \midrule
      \midrule
\multicolumn{1}{c}{Noun}&\multicolumn{1}{c}{Occurrences}&\multicolumn{1}{c}{Context}\\
      \midrule
presentation&164&[``lucid", 0.7, ``JJ", ``the presentation is lucid and provided examples"]\\
essay&88&[``balanced", 0.9, ``JJ", ``was a very balanced essay."]\\
topic&60&[``relevant", 0.5, ``JJ", ``clearly explained the relevant topic briefly."]\\
team&54&[``successful", 0.9, ``JJ", ``team has been successful in capturing all"]\\
animation&10&[``impressive", 1.0, ``JJ", ``it was really impressive and the animation"]\\
      \midrule
      \midrule
   \end{tabular}
 }
\end{table*}

\subsection{Lexicon Check}

\begin{figure}[h]
\centering
\includegraphics[scale=0.51]{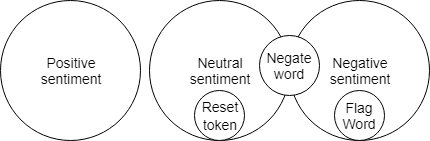}
\caption{Polarity of tokens.}
\label{fig:dict}
\end{figure}

\begin{figure}[h]
\centering
\includegraphics[scale=0.51]{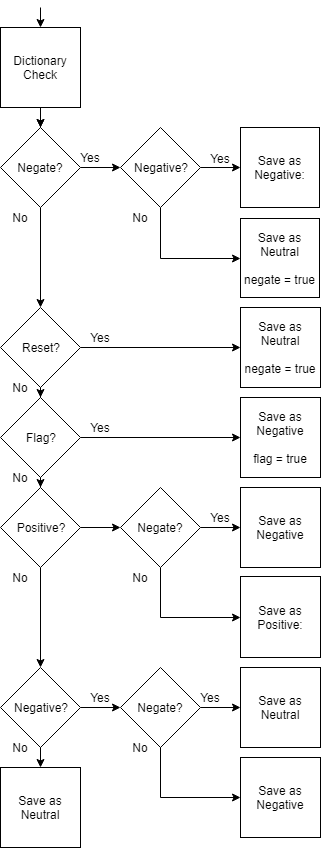}
\caption{Flowchart for the lexicon check.}
\label{fig:dictCheck}
\end{figure}

To classify the polarity of words, we utilized tokens (words and punctuation) from six categories (\autoref{fig:dict}): positive sentiment, negative sentiment, neutral sentiment, negate word, flag word, and reset token. There was overlap between negative sentiment and negate words (e.g., ``missing") as well as between negative sentiment and flag words (e.g., ``copying"). Reset tokens were always neutral (e.g., ``however"). We processed tokens in a specific order (\autoref{fig:dictCheck}) and saved tokens as either neutral, positive, or negative.

\subsection{Negation}

%Any language includes complicated sentence structure, irony, sarcasm, or any number of things that can alter the tone of a comment, and English is no exception. Thus, one key aspect of our sentiment algorithm is capturing negated sentiment

Sentiment negating words, or ``valence shifters," were perhaps the most complicated and interesting component of the sentiment grader and they occurred in just over 47\% of our reviews. Words in our negate dictionary were comprised of regular negation words (e.g., ``none''), presuppositional words (e.g., ``barely''), omission words (e.g., ``missing''), and modal auxiliary verbs (e.g., ``should''). Negation is typically associated with negative sentiment, not distributed equally with positive sentiment \cite{potts2010negativity}, a finding we also observe in \autoref{analysis}.\textremoved{This section includes actual examples captured and correctly classified.} Similarly to  \cite{wiegand2010survey}, we found negating negative sentiment changed tone to neutral (e.g., ``it wasn't terrible" or ``not wordy") but negating positive sentiment changed tone to negative (e.g., ``it wasn't clear" or ``nothing innovative"). Our process captured negated positive words in three ways (see \cite{taboada2011lexicon} for alternatives):

%It is important to note that, in general, negating negative sentiment changes tone to neutral (e.g., ``it wasn't terrible" or ``not wordy") but negating positive sentiment changes tone to negative (e.g., ``it wasn't clear" or ``nothing innovative"). Our process captured negated positive words in three ways (see \cite{taboada2011lexicon} for alternatives):

\begin{itemize}[noitemsep]
\item negate word to reset token;
\item preceding negative qualifier;
\item trailing negative qualifier.
\end{itemize}

Certain words negated sentiment until their meaning was removed with a reset token (e.g., \cite{pang2002thumbs}):
\begin{quote} 
\centering 
``...could have given \textit{\textbf{more} practical examples to make it more clear as several readings were required to understand the topic and their idea flow}."
\end{quote}
The positive words/phrases: practical examples, clear, understand the topic, and idea flow were all negated until the reset token, `.', was encountered. Other reset tokens that conclude a negate word's effect include: `.', `;', but, although, however, and nevertheless. Negate words include: no, not, can't, nothing, hardly, barely, lack, more, suggest, miss, and few.

Secondly, polarity was negated through a preceding negative qualifier. Simply put: a negative adjective in close proximity before a positive word (``close" is intentionally vague and is configurable in the algorithm):
\begin{quote} 
\centering 
``...which makes it \textit{\textbf{difficult} to understand}..."
\end{quote}

%To `understand' is a positive sentiment, but the near negative sentiment reverses its polarity.
Finally, positive tone was negated through a trailing negative qualifier. This required observation into the future as the algorithm scanned the sliding window of text for a negative adjective which negated positive sentiment:

\begin{quote} 
\centering 
``...some \textit{insight was \textbf{missing}}..."
\end{quote}

It is important to note that the preceding/trailing negative qualifier was itself a negative sentiment word, thus ``stacking" the effect of its negativity. This was intentional, as adjectives are often sentiment magnifiers \cite{neviarouskaya2011sentiful}, \cite{wilson2005recognizing}. In contrast, negate words could either be negative or neutral, as seen in the overlap of the negate word lexicon in \autoref{fig:dict}.

\subsection{Grading} \label{grade}
\begin{figure}[h]
\centering
\includegraphics[scale=0.51]{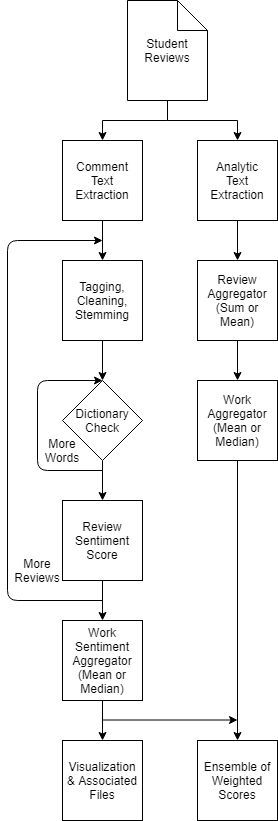}
\caption{Flowchart for the grading process.}
\label{fig:grade}
\end{figure}

Our grading process (\autoref{fig:grade}) extracted the analytical and subjective content into two files, which were processed by two different graders. The analytic grader simply matched student responses with their assigned values and aggregated the score. The process was relatively straightforward and is detailed in a prior paper \cite{beasley2019designing}.\textinserted{Ultimately, the sentiment score (with accompanying metrics) was weighted with the analytic score and provided to the instructor as a suggested final grade from the peer-review process.} 

%\subsection{Metrics \& Attributes}

We utilized a variety of metrics and attributes to provide the maximum information to an instructor. Some are widely used in sentiment analysis (e.g., `tone' and `purity'), while others were specific to our process (e.g., `dif'). The following metrics contributed to a sentiment score \textit{per review comment}:

\begin{itemize}[noitemsep]
\item weight: the sentiment of a lexicon-matched keyword: [0, 1];
\item pos\_keywords: the number of positive keywords: 0 to inf;
\item neg\_keywords: the number of negative keywords: 0 to inf;
\item keywords: the total number of keywords matched: 0 to inf;
\item tone: the sum of all weighted keywords: (-inf, inf);
\item info: the absolute value of all weighted keywords: [0, inf);
\item score: the sentiment score (F to A+): [0, 4.3].
%\item score: the sentiment score (scaled C to A+): [1.8, 4.3]
\end{itemize}

The following attributes were defined or derived for validation and analysis \textit{per review comment}:
\begin{itemize}[noitemsep]
%\item reliability: `true' if there is a score, and info greater than 2 or keywords greater than 4: 0 or 1
\item reliable: whether a comment has enough information to be aggregated: 0 or 1;
\item default: whether a comment matched a threshold of keywords and was scored: 0 or 1;
%\item dif: the difference between sentiment score and analytic section score: [-2.5, 4.3]
\item dif: the difference between sentiment score and analytic section score: [-4.3, 4.3];
\item purity: tone over info, a measure of how consistent the sentiment is: [-1, 1];
\item positivity: the sum of positive sentiment: [0, inf);
\item negativity: the sum of negative sentiment: (-inf, 0];
\item negate\_words: the number of negating words: 0 to  inf;
\item words\_per\_sentence: the number of words per sentence: 0 to inf;
\item length: the number of words in the review: 0 to inf.
\end{itemize}

The following attributes were defined for validation and analysis \textit{for an aggregation of peer reviews}:
\begin{itemize}[noitemsep]
\item pos\_dict\_used: the percentage of tokens in the positive lexicon utilized: [0, 1];
\item neg\_dict\_used: the percentage of tokens in the negative lexicon utilized: [0, 1].
\end{itemize}

\begin{figure}[h]
\centering
\includegraphics[scale=0.58]{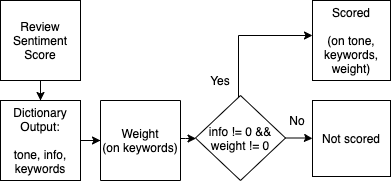}
\caption{Flowchart for the sentiment scoring algorithm.}
\label{fig:score}
\end{figure}

Most importantly, our lexicon check for a single review comment provided \textit{tone} and \textit{keywords}. We assigned a fine-grained score to the text (\autoref{fig:score}) by \textit{tone over keywords} with some adjustments including scaling to the range desired by the instructor. Thus the highest scores were a result of many highly weighted positive keywords (i.e., highly positive purity), and lowest scores were those with many highly negative keywords (i.e., highly negative purity). We tested two options for aggregating comments by median and mean:

\begin{enumerate}[noitemsep]
  \item weight reviews based on information available (complex);
  \item weight all reviews equally (simple).
\end{enumerate}

\begin{figure}[h]
\centering
\includegraphics[scale=0.57]{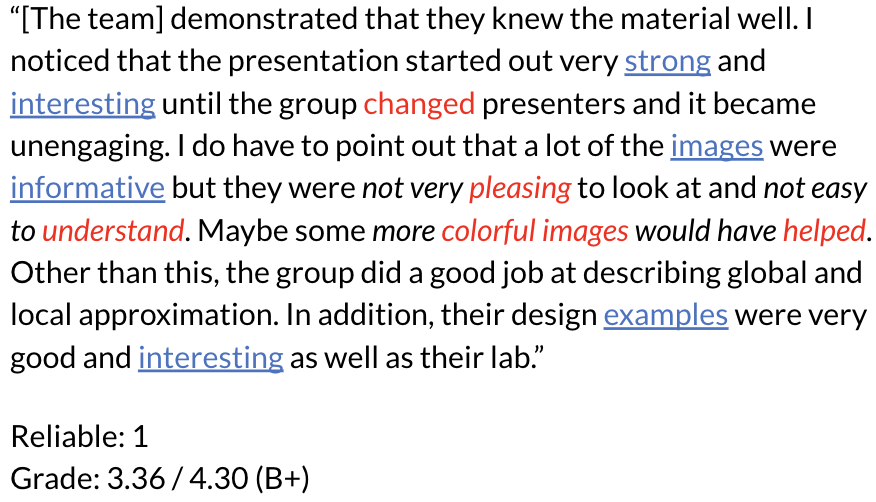}
\caption{Example of the sentiment scoring algorithm applied to a student's review. Net positive is blue and underlined, net negative is red, and negated sentiment is italicized.}
\label{fig:sentiScore}
\end{figure}

In the first grading scheme, we began by weighting our confidence in the comment. If the sentiment was negative and lacked many keywords, we reduced the weight significantly to avoid penalizing harshly. If the review was positive with little information, we weighted slightly less to compensate. The second grading scheme in effect tested the principle of the wisdom of the crowd and allowed all reviews, even ones with a single keyword, to contribute to a student's score. This required a leap of faith---trusting that a few reviews with highly negative sentiment would not destroy a student's grade. In either scheme, if there was not enough information to process the comment we simply incremented the number of default scores. Over the last five semesters, roughly 75\% of our reviews meet our basic threshold for scoring with weighted confidence.\textinserted{This number increased to approximately 85\% when we incentivized students to provide a high-quality comment with a completion point. Reviews were checked by the teaching assistant and students quickly adjusted to providing in-depth reviews.} \autoref{fig:sentiScore} is an example of the sentiment grader on a single student's review (net positive -- blue and underlined, net negative -- red, negated -- italicized), with the score and reliability generated from the complex scorer.

%If a review fails to meet a basic criteria for sentiment, it cannot be scored confidently and will not contribute to a student's score. 

%We modified the scorer to simply return tone over keywords if there was any information in the review.
 %for our two most recent courses where students were incentivized to provide quality review comments: Geometric Modeling (GM), a graduate-level course, and Software Engineering (SE), an undergraduate course.\\ 

%Finally, we aggregated the sentiment scores by median and mean. The sentiment score could then be weighted with other review form categories into a student's final score. 

%Unsurprisingly, when we simplified to the second grading scheme for our two most recent courses where students were incentivized to provide quality review comments, the means decreased on average: $ \sim $0.18 (4.2\%) for Software Engineering (SE), an undergraduate course, and $ \sim $0.227 (5.3\%) for Geometric Modeling (GM), a graduate course. The medians were also affected, although not as severely. If we did a simple boost of 0.125 (2.9\%) on the means of each review, the mean was reduced by only $ \sim $0.021, or 0.5\% for SE and $ \sim $0.071, or 1.7\% for GM. \autoref{fig:gmScore} and \autoref{fig:seScore} show the difference in means using both scorers. One reason we saw some jumps between the grading schemes was that reviews with fewer keywords (especially negative reviews) were previously weighted significantly lower (or even disregarded) so they would not sway the final score. A side effect of this simpler grading scheme was that our average percentage of reliable reviews went down (along with our positive sentiment and and average words because we kept terser, more negative reviews).

\begin{figure}[h]
    \centering
    \begin{subfigure}[t]{0.5\textwidth}
        \centering
  \includegraphics[width=.98\linewidth]{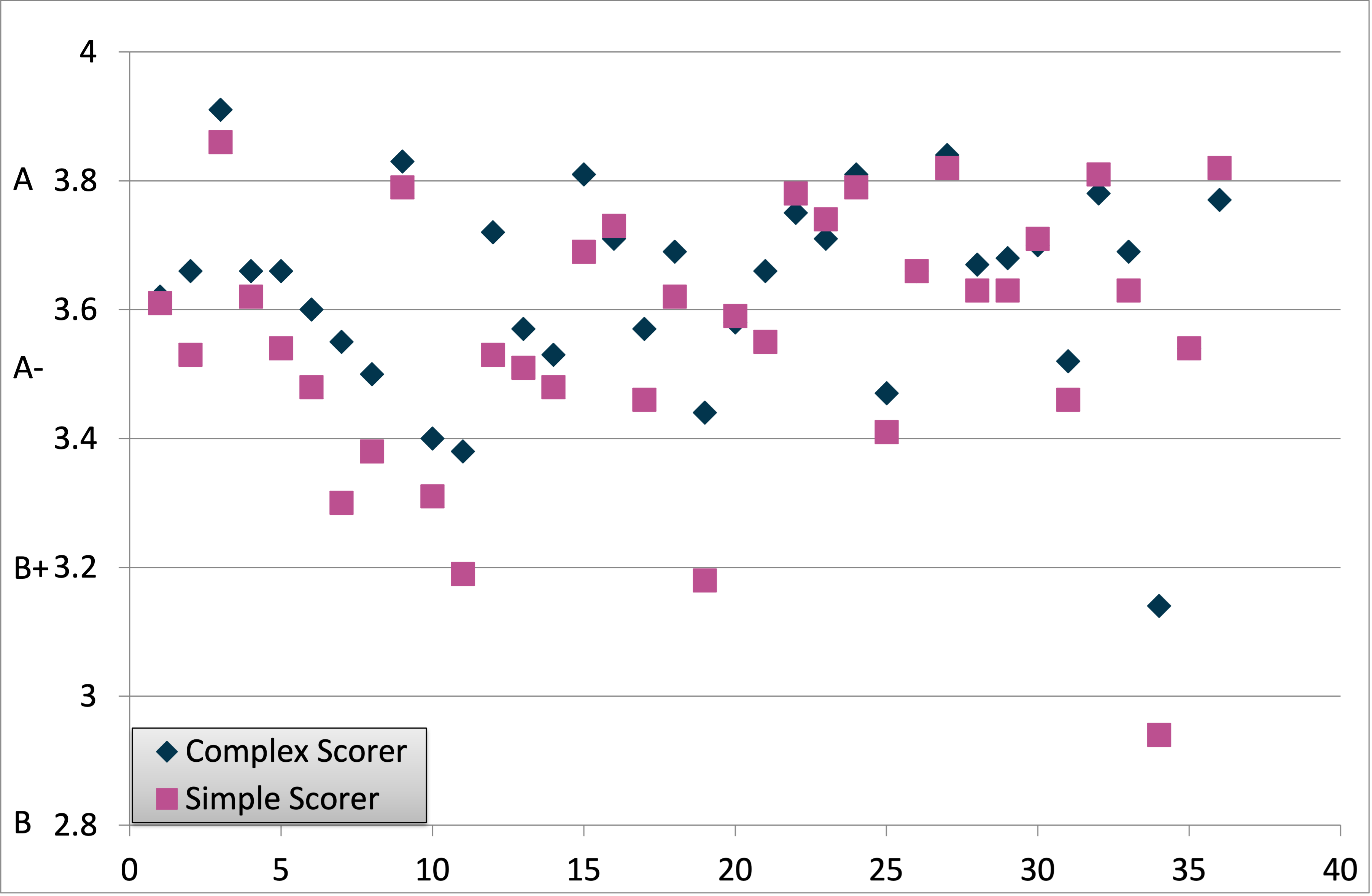}
\caption{GM (average difference -0.068)}
\label{fig:gmScore}
\vspace*{5mm}
    \end{subfigure}
    ~ 
    \begin{subfigure}[t]{0.5\textwidth}
        \centering
\includegraphics[width=.98\linewidth]{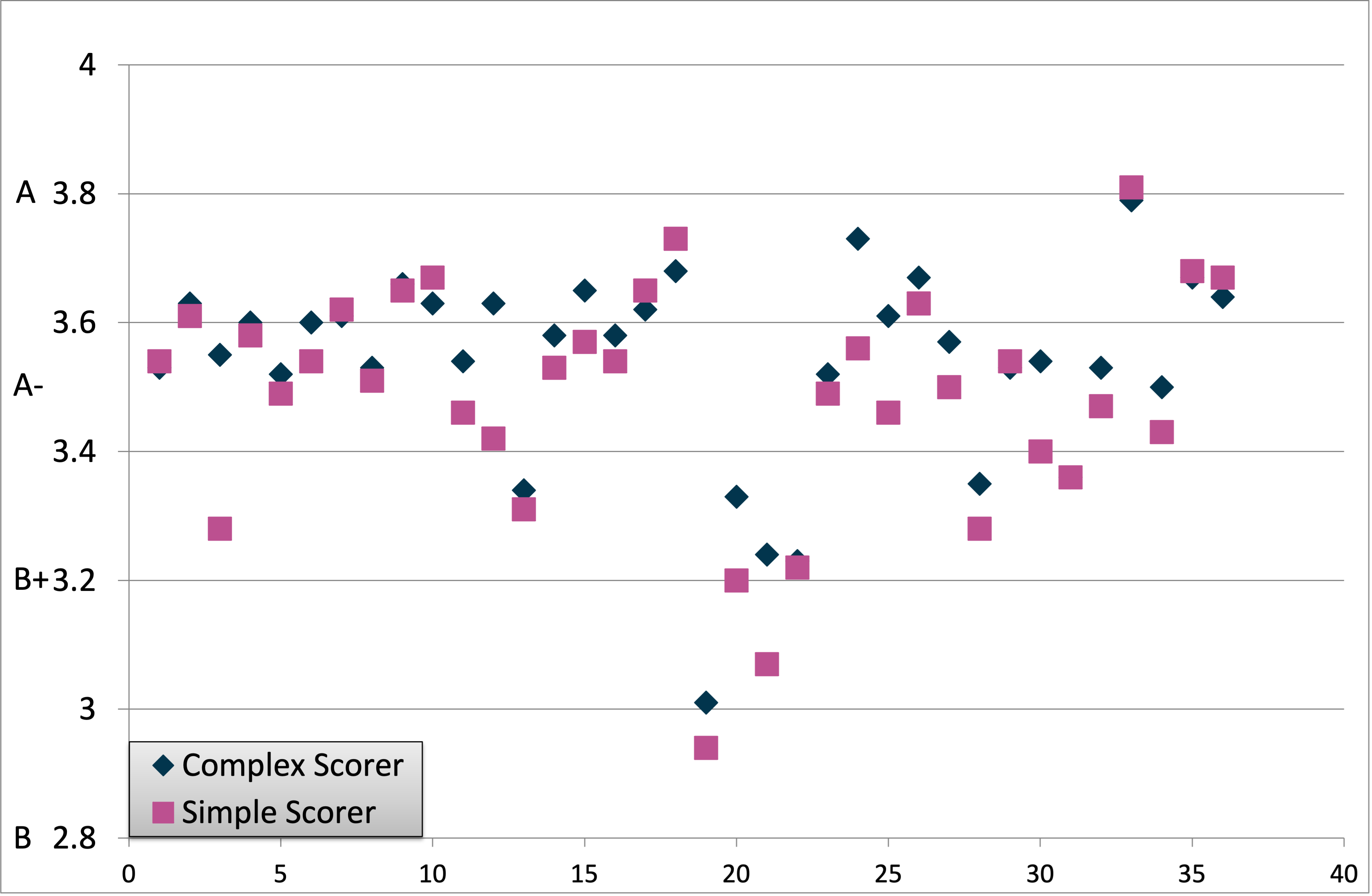}
\caption{SE (average difference -0.053)}
\label{fig:seScore}
    \end{subfigure}
\caption{Change in (a) Geometric Modeling and (b) Software Engineering mean project scores (y-axis) using different scoring algorithms for each review (x-axis).}
\label{fig:scores}
\end{figure}

When we simplified to the second grading scheme for our two most recent courses where students were incentivized to provide quality review comments, the means decreased on average, although the difference was minute: $ \sim $0.053 (1.2\%) for Software Engineering (SE), an undergraduate course, and $ \sim $0.068 (1.6\%) for Geometric Modeling (GM), a graduate course. The medians were also affected, although less notably (0.68\% and 0.15\%, respectively). \autoref{fig:gmScore} and \autoref{fig:seScore} show the difference in means using both scorers. One reason we saw some jumps between the grading schemes was that reviews with fewer keywords (especially negative reviews) were previously weighted significantly lower (or even disregarded) so they would not sway the final score. A side effect of this simpler grading scheme was that our average percentage of reliable reviews went down along with our positive sentiment and average words because we kept terser, more negative reviews.

%\subsubsection{Simple Scorer}

Simplifying the grader also had the side effect of increasing the average standard deviation of SE by 2.00\% to 0.435 and GM by 3.07\% to 0.479. Although the average standard deviation was slightly high and above our target of 10\% of the grade range, or 0.430, we relied on the principle of the wisdom of the crowd to postulate that the means were still accurate. The fact that major algorithm changes (especially simplifications) only shifted grades within a fraction of a letter grade ($<$ 2\%) suggested that we had a robust sentiment analysis algorithm when coupled with a large number of peer reviews per work.
% that can be trusted to reliably provide a sentiment grade.

\begin{comment}
% This doesn't have the new pdfs
\begin{figure}[h]
\centering
\includegraphics[scale=0.4]{gm_score_new.png}
\caption{GM Change in Score (average difference -0.071)}
\label{fig:gmScore}
\end{figure}

\begin{figure}[h]
\centering
\includegraphics[scale=0.4]{se_score_new.png}
\caption{SE Change in Score (average difference -0.021)}
\label{fig:seScore}
\end{figure}
\end{comment}

\section{Validation} \label{valid}

Firstly, sentiment score roughly tracked the review form score. In SE and GM, the review form score was slightly higher on average than sentiment mean (0.323 and 0.121 in the weighted versions and 0.376 and 0.190 in the unweighted versions). This ranged from 2.82\% to 8.73\% of total score.\textinserted{We did note some deviation between the \textit{individual} scores---a phenomenon also observed in other domains \cite{dave2003mining}, \cite{fang2015sentiment}. This deviation suggested that students were willing to write honestly, even if they gave their peers a good grade via the analytical section. Unfortunately, this fact prevented us from using the review form analytical score as a ground truth against the sentiment score. It also precluded us from using an algorithm from another domain, since we would have no way of quantitatively verifying the sentiment score.}

\textinserted{However, in a prior work \cite{beasley2021domain}, we quantitatively compared the mean average error (MAE) between the \textit{aggregate} review form analytic score and \textit{aggregate} review form sentiment score for six publicly available lexicons: Affective Norms for English Words (ANEW) \cite{bradley1999affective}, SlangSD \cite{wu2016slangsd}, Multi-Perspective Question Answering (MPQA 3.0) \cite{deng2015mpqa},  Valence Aware Dictionary and sEntiment Reasoner (Vader) \cite{gilbert2014vader}, SentiWordNet 3.0 \cite{baccianella2010sentiwordnet}, and AFINN-111 \cite{afinn} while holding the scoring algorithm constant. This work showed that our lexicon captured roughly the same \textit{amount} of sentiment even though the other lexicons were much larger \cite{beasley2021domain}.} GM and SE reviews carried enough information to be counted. We average 4.24 positive keywords, 1.48 negative keywords, and 1.13 negate words per review. Percent default scores ranged from 0\% to 13\% (3.6\% on average for CG and 2.1\% for SE) with 63\% of reviews marked as reliable for the simplified scorer.

\begin{figure*}[h]
\centering
\fbox{\includegraphics[scale=0.62]{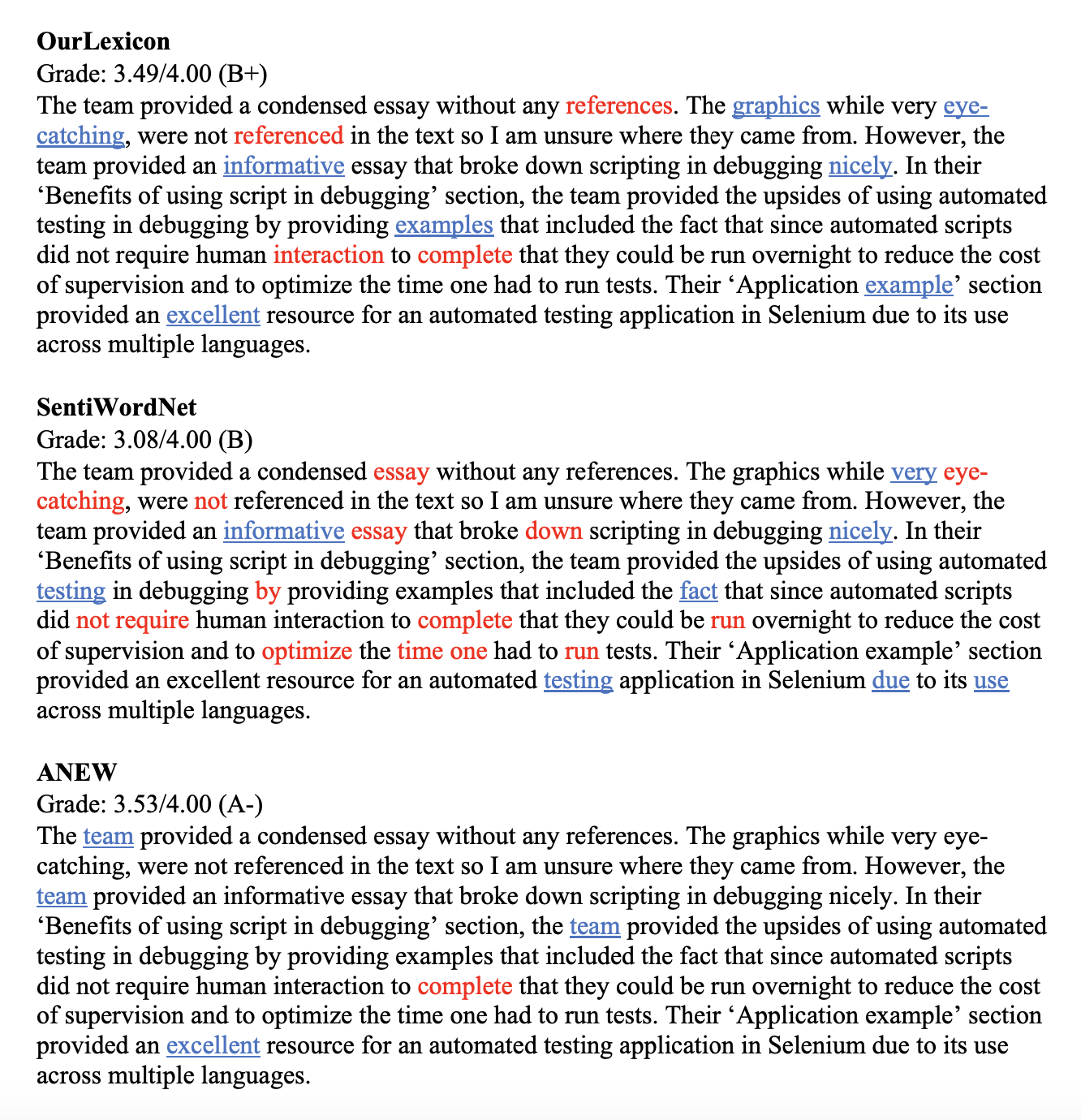}}
\caption{Qualitative comparison of three lexicons. Net positive is blue and underlined, net negative is red.}
\label{fig:lex_comp}
\end{figure*}

\textinserted{Additionally, the \textit{quality} of the sentiment collected was the most precise from our lexicon. \autoref{fig:lex_comp} is one qualitative example that demonstrates the difference in how the top three sentiment-producing lexicons---ours, ANEW, and SentiWordNet---interpreted text from the same review. This demonstrates each lexicon's capability of ``understanding" the text by correctly noting sentiment. The closer the interpretation of positive and negative sentiment matches human interpretation, the better the lexicon. SentiWordNet captured the most sentiment, but not in an intuitive way, labeling words like ``essay," ``eye-catching," and ``down" as negative or ``due" and ``use" as positive. ANEW, with keywords selected and weighted by hand, matched words more intuitively, but did not capture clear sentiment-bearing words like ``informative" or ``nicely." ANEW also coded ``team" as positive rather than neutral. In contrast, our lexicon correctly identified a large quantity of keywords and appropriately tagged the quality of sentiment while ignoring neutral words. It captured both the specific examples provided by the team and the lack of references as well as the positive keywords ``excellent" and ``informative."}

Finally, the infrequency of instructor adjustment to the algorithm score was evidence for its correctness.\textremoved{There are three options to handle differences between crowd and instructor score: 1) accept the wisdom of the crowd after we have established the algorithm and the minimum number of reviewers, 2) use the crowd score to adjust the grade of the instructor, or 3) adjust the crowd score.} In five semesters of use (nine courses, 324 student submissions), the instructor only changed the algorithm's score \textit{once}. In this situation, the score was adjusted up to compensate for a number of students reporting a failing score for ``reading off the slides''. This problem was further compounded by the rather coarse-grained overall score selection (scale of four instead of the current eight) from a previous review form iteration. Increasing the granularity of overall score softened the otherwise large variation. Ultimately, testing different grading methods with our lexicon and finding they produce scores that are almost identical is an indication of a statistically sound procedure \cite{hoffmann2019benchmarking}).

\section{Discussion} \label{analysis}

We evaluated the Pearson correlation coefficient on each metric from SE and GM with the mean and median simplified sentiment scores and used a significance of correlation test (degrees of freedom $(df) = 70$ and critical value of $+/- .380$). \autoref{table:correlations} shows a selection of correlation values ($r$) that are statistically significant at the $\alpha = .001$ level. Mean and median scores correlated very closely (.935), suggesting that either may be utilized as the final score and that we have an approximately normal distribution. Purity was also highly correlated with mean which provided our first indication that we could simplify the scoring algorithm.

% Surprisingly, the review form score (30\% of which is based off of the sentiment score) is comparatively loosely correlated with sentiment score. This suggests that the sentiment score adds some new information not contained in the review form itself, which bolsters the importance of its use.\\

% STATISTICALLY SIGNIFICANT CORRELATIONS WITH MEAN AND MEDIAN
\begin{table*}[!t]
\centering
\caption{\\Statistically Significant Correlations with Mean and Median}
\label{table:correlations}
\resizebox{2\columnwidth}{!}{%
   \begin{tabular}{@{} l*{9}{l} @{}}
      \midrule
      \midrule
&FormScore&StdDev&NegSenti&Senti&Purity&NegKey&NegateKey&Adverbs\\
      \midrule
Mean&.750&-.462&.830&.764&.982&-.814&-.829&-.373\\
Median&.696&-.281&.820&.696&.913&-.821&-.813&-.432\\
      \midrule
      \midrule
   \end{tabular}
 }
\end{table*}

%0.232 is the critical value for p < 0.05 for 70 df

To discover what information is added from the simple sentiment scorer, we analyzed the correlation of mean with various metrics to infer the following: 
\begin{enumerate}[noitemsep]
  \item Negate keywords have a similar negative correlation with the mean compared to Negative Words. This suggests that the role of negating words factors highly into grading, and situations in which they occur should be carefully handled;
  \item Standard Deviation is negatively correlated with the mean. This suggests that most reviews are positive and that increased deviation comes from additional, negative reviews;
  \item Percent Reliable has a very slight negative correlation with the mean (-.127). This suggests that most reliable reviews are negative and most unreliable reviews are positive;
  \item Finally, a number of length and keyword metrics are negatively correlated with the mean (significant at the $\alpha = .05$ critical value of $+/- .232$): Total Keywords (-.249), Words (-.284), Words/Sentence (-.258), and especially Adverbs (-.373). This suggests that \textit{the more a student writes, the more faults they find in others' works} (even though just 15\% of all reviews contain more negative sentiment than positive sentiment).
\end{enumerate}

Taken together, this information indicates that many students write cursory, unreliable positive reviews, but a few diligent students write longer, more detailed, and more negative reviews. We find such negative reviews more helpful in an academic context, and would be interested in analyzing whether those reviewers grade more like an instructor or are higher performing students. These reviews can be filtered to present to the instructor a more balanced and realistic perspective of the students' works.

\textinserted{We also analyzed the top one, five, ten, fifteen, and twenty percent of the most positive and most negative reviews to determine the top three keywords (including ties) most represented (\autoref{table:topKeywords}). GM had 920 reviews, while SE had 1027. It should be noted that only positive or negative keywords are included, with negated text excluded. For example, the following student's review is negative but only contains negated positive sentiment and not negative keywords: ``They took a safe road quite literally. The safe itself looks standard and nothing like high end or unique but ifs [sic] a safe. They could have added more features to make it unique."

On the negative side, ``hard," ``need," and ``short" quickly saturated the most negative reviews in both courses. On the positive end of the spectrum, ``impressive" and ``unique" seemed important to the reviewers in GM (who were designing models), while SE students (who were creating software systems) highlighted ``useful" work that was easy to ``understand." Students in both courses appreciated the inclusion of ``examples." This type of insight can be utilized in courses even where a different peer-review format is utilized. In fact, our process, specifically aspect extraction and textual metrics, was utilized to deliver insights in three Data Visualization courses on factors related to student engagement in and enjoyment of peer review \cite{beasley2020leveraging}.}

% GM, SE 2019
% KEYWORDS IN MOST POSITIVE AND NEGATIVE REVIEWS
\begin{table*}[!t]
\centering
\caption{\\Keywords in the Most Positive and Negative Reviews}
\label{table:topKeywords}
\resizebox{2\columnwidth}{!}{%
   \begin{tabular}{@{} l*{3}{l} @{}}
      \midrule
      \midrule
%&\multicolumn{2}{c}{GM}&\multicolumn{2}{c}{SE}\\
%\hline
%&\multicolumn{1}{c}{Positive}&\multicolumn{1}{c}{Negative}&\multicolumn{1}{c}{Positive}&\multicolumn{1}{c}{Negative}\\
&\multicolumn{1}{c}{Most Positive}&\multicolumn{1}{c}{Most Negative}\\
      \midrule
GM 1\%&unique(3), creative(3), outstand(3)&need(1)\\
GM 5\%&unique(15), outstand(11), creative(9)&hard(3), need(3), short(3)\\
GM 10\%&unique(25), creative(19), structure(18), example(18)&short(14), hard(7), need(5)\\
GM 15\%&example(37), unique(37), impressive(32)&hard(16), need(15), short(15)\\
GM 20\%&impressive(60), example(57), unique(54)&hard(21), need(21), short(21)\\
&&\\
SE 1\%&fantastic(3), engage(2), enjoy(2), useful(2), outstand(2)&punctuation(1), incorrect(1), awkward(1), heavy(1), bor(1), stuck(1)\\
SE 5\%&useful(15), outstand(10), enjoy(9), unique(9), fantastic(9)&need(10), short(6), add(3), bor(3), heavy(3), hard(3)\\
SE 10\%&useful(25), example(25), enjoy(18), outstand(18)&need(21), short(16), hard(11)\\
SE 15\%&example(52), useful(32), outstand(30)&need(36), short(24), hard(15)\\
SE 20\%&example(87), understand(40), useful(40)&need(48), short(30), hard(23)\\
      \midrule
      \midrule
   \end{tabular}
 }
\end{table*}

\section{Conclusion}
\textinserted{Sentiment analysis is uniquely suited to increase and summarize the information gathered from peer-review text in large courses, which allows greater confidence in accurately grading open-ended assignments. Applying natural language processing techniques to the data generated from an intelligent, reviewing crowd increases the ability to capture content less accessible to automated graders: humor, beauty, sarcasm, or truthfulness. A lexicon-based approach increases information and allows for aspect extraction, which semiautomates review form modification based on features in actual student reviews. By utilizing our generalizable \textit{process}, an instructor can approach assessment through crowdsourced peer review in a data-driven way. Although our approach was created in the context of a computer science course, mining information from peer-review text can yield a system tuned to any course's context. Ultimately, extracting sentiment from peer-review text is an untapped area with the potential to increase information and reliability in assessment.}

\textremoved{We present an application of peer review sentiment analysis toward accurately scaling grading in large, online educational systems. Leveraging a lexicon-based approach allows for aspect extraction, which semiautomates review form modification based on features in actual student reviews. Rather than providing a one size fits all review form or sentiment lexicon, we present a generalizable process by which an instructor can approach assessment through crowdsourced peer review in a data-driven way.}

%We present an application of peer review sentiment analysis towards accurately scaling grading in large, online educational systems. Leveraging a lexicon-based approach allows for aspect extraction, which semiautomates review form modification based on features in actual student reviews. Rather than providing a one size fits all review form or sentiment lexicon, we present a generalizable process by which an instructor can approach assessment through crowd-sourced peer review in a data-driven way.\textinserted{Although our approach was created in a computer science course context, mining information from peer review text will yield a system tuned to any course's context. Elsewhere we confirm the application of our process, namely aspect extraction and textual metrics, to deliver insights in courses where a different peer review format is utilized \cite{beasley2020leveraging}. Ultimately, extracting sentiment from peer review text is an untapped area with enormous potential for increasing information and reliability in assessment.}

In the future, we would like to better utilize the aspect generator via double propagation to use 1) known aspects to find other aspects, 2) known aspects to find other sentiment words, and 3) known sentiment words to find other sentiment words via connectors (e.g., ``and'', ``but'', ``however''). We are also interested in adopting a visualization to condense all peer reviews into an understandable graphic (perhaps using aspect-level opinion mining)\textremoved{along with 1--3 most representative reviews}to increase the information presented to an instructor for grading confidence.
% opinion mining similar to \cite{liu2015seeing}

%In the future, we would like to visualize the difference between reviews scored with an open-sourced recurrent neural network (trained on comments labeled by hand or from Amazon Mechanical Turk) and our simplified algorithm. We would also like to extend our review form analysis to non-engineering courses, to increase confidence in the scope of our process to different sub-domains within academia. Furthermore, we can better utilize the aspect generator via double propagation to use 1) known aspects to find other aspects, 2) known aspects to find other sentiment words, and 3) known sentiment words to find other sentiment words via connectors (e.g., ``and'', ``but'', ``however''). Finally, we are interested in adopting a visualization to condense all peer reviews into an understandable graphic (perhaps using aspect-level opinion mining similar to \cite{liu2015seeing}) to increase the information presented to an instructor for grading confidence. With this visualization, we would like to present to an instructor 1-3 most representative reviews.

\section{ACKNOWLEDGMENT}
We thank our reviewers for their helpful feedback on our paper. The project is supported in part by the National Science Foundation (IIS-1845204).

\begin{comment}

% if have a single appendix:
%\appendix[Proof of the Zonklar Equations]
% or
%\appendix  % for no appendix heading
% do not use \section anymore after \appendix, only \section*
% is possibly needed

% use appendices with more than one appendix
% then use \section to start each appendix
% you must declare a \section before using any
% \subsection or using \label (\appendices by itself
% starts a section numbered zero.)
%

\appendices
\section{Proof of the First Zonklar Equation}
Appendix one text goes here.

% you can choose not to have a title for an appendix
% if you want by leaving the argument blank
\section{}
Appendix two text goes here.

% use section* for acknowledgment
\section*{Acknowledgment}

The authors would like to thank...
\end{comment}

% Can use something like this to put references on a page
% by themselves when using endfloat and the captionsoff option.
\ifCLASSOPTIONcaptionsoff
  \newpage
\fi

% trigger a \newpage just before the given reference
% number - used to balance the columns on the last page
% adjust value as needed - may need to be readjusted if
% the document is modified later
%\IEEEtriggeratref{8}
% The "triggered" command can be changed if desired:
%\IEEEtriggercmd{\enlargethispage{-5in}}

% references section

% can use a bibliography generated by BibTeX as a .bbl file
% BibTeX documentation can be easily obtained at:
% http://mirror.ctan.org/biblio/bibtex/contrib/doc/
% The IEEEtran BibTeX style support page is at:
% http://www.michaelshell.org/tex/ieeetran/bibtex/
\bibliographystyle{IEEEtran}
% argument is your BibTeX string definitions and bibliography database(s)
%\bibliography{IEEEabrv,../bib/paper}
%
% <OR> manually copy in the resultant .bbl file
% set second argument of \begin to the number of references
% (used to reserve space for the reference number labels box)
\bibliography{Polarity}

% Generated by IEEEtran.bst, version: 1.12 (2007/01/11)
\begin{thebibliography}{10}
\providecommand{\url}[1]{#1}
\csname url@samestyle\endcsname
\providecommand{\newblock}{\relax}
\providecommand{\bibinfo}[2]{#2}
\providecommand{\BIBentrySTDinterwordspacing}{\spaceskip=0pt\relax}
\providecommand{\BIBentryALTinterwordstretchfactor}{4}
\providecommand{\BIBentryALTinterwordspacing}{\spaceskip=\fontdimen2\font plus
\BIBentryALTinterwordstretchfactor\fontdimen3\font minus
  \fontdimen4\font\relax}
\providecommand{\BIBforeignlanguage}[2]{{%
\expandafter\ifx\csname l@#1\endcsname\relax
\typeout{** WARNING: IEEEtran.bst: No hyphenation pattern has been}%
\typeout{** loaded for the language `#1'. Using the pattern for}%
\typeout{** the default language instead.}%
\else
\language=\csname l@#1\endcsname
\fi
#2}}
\providecommand{\BIBdecl}{\relax}
\BIBdecl

\bibitem{shah2014some}
N.~B. Shah, J.~Bradley, S.~Balakrishnan, A.~Parekh, K.~Ramchandran, and M.~J.
  Wainwright, ``Some scaling laws for {MOOC} assessments,'' in \emph{KDD
  Workshop Data Mining for Educational Assessment and Feedback (ASSESS '14)},
  New York City, NY, USA, 2014.

\bibitem{singh2016question}
G.~Singh, S.~Srikant, and V.~Aggarwal, ``Question independent grading using
  machine learning: The case of computer program grading,'' in \emph{Proc. 22nd
  ACM Int. Conf. Knowledge Discovery and Data Mining (SIGKDD '16)}, San
  Francisco, CA, USA, Aug. 13, 2016, pp. 263--272, doi:
  \url{10.1145/2939672.2939696}.

\bibitem{kulkarni2014scaling}
C.~E. Kulkarni, R.~Socher, M.~S. Bernstein, and S.~R. Klemmer, ``Scaling
  short-answer grading by combining peer assessment with algorithmic scoring,''
  in \emph{Proc. 1st ACM Conf. Learning at Scale (L@S '14)}, Atlanta, GA, USA,
  Mar. 4, 2014, pp. 99--108, doi: \url{10.1145/2556325.2566238}.

\bibitem{beasley2018ten}
Z.~J. Beasley, L.~A. Piegl, and P.~Rosen, ``Ten challenges in cad cyber
  education,'' \emph{Comput.-Aided Des. and Appl.}, vol.~15, no.~3, pp.
  432--442, May 2018, doi: \url{10.1080/16864360.2017.1397893}.

\bibitem{beasley2019designing}
------, ``Board 39: Designing intelligent review forms for peer assessment: A
  data-driven approach,'' in \emph{Proc. 2019 ASEE Annu. Conf. \& Expo.},
  Tampa, FL, USA, Jun. 15, 2019, doi: \url{10.18260/1-2--32337}.

\bibitem{russell2017variability}
J.~Russell, S.~Van~Horne, A.~Ward, E.~Bettis~III, and J.~Gikonyo, ``Variability
  in students' evaluating processes in peer assessment with calibrated peer
  review,'' \emph{J. Comput. Assisted Learn.}, vol.~33, no.~2, pp. 178--190,
  Apr. 2017, doi: \url{10.1111/jcal.12176}.

\bibitem{almatrafi2018systematic}
O.~Almatrafi and A.~Johri, ``Systematic review of discussion forums in massive
  open online courses ({MOOCs}),'' \emph{IEEE Trans. Learn. Technol.}, Jul.
  2018.

\bibitem{li2019does}
H.~Li, Y.~Xiong, C.~V. Hunter, X.~Guo, and R.~Tywoniw, ``Does peer assessment
  promote student learning? a meta-analysis,'' \emph{Assessment \& Eval. Higher
  Educ.}, pp. 1--19, Feb. 2019, doi: \url{10.1080/02602938.2019.1620679}.

\bibitem{somasundaran2018towards}
S.~Somasundaran, M.~Flor, M.~Chodorow, H.~Molloy, B.~Gyawali, and L.~McCulla,
  ``Towards evaluating narrative quality in student writing,'' \emph{Trans.
  Assoc. Comput. Linguistics}, vol.~6, pp. 91--106, Jan. 2018, doi:
  \url{10.1162/tacl_a_00007}.

\bibitem{moyne2018development}
M.~M. Moyne, M.~Herman, K.~Z. Gajos, C.~J. Walsh, and D.~P. Holland, ``The
  development and evaluation of {DEFT}, a web-based tool for engineering design
  education,'' \emph{IEEE Trans. Learn. Technol.}, vol.~11, no.~4, pp.
  545--550, Feb. 2018, doi: \url{10.1109/TLT.2018.2810197}.

\bibitem{liu2012sentiment}
B.~Liu, ``Sentiment analysis and opinion mining,'' \emph{Synthesis Lectures
  Human Lang. Technol.}, vol.~5, no.~1, pp. 1--167, May 2012, doi:
  \url{10.2200/S00416ED1V01Y201204HLT016}.

\bibitem{serrano2015sentiment}
J.~Serrano-Guerrero, J.~A. Olivas, F.~P. Romero, and E.~Herrera-Viedma,
  ``Sentiment analysis: A review and comparative analysis of web services,''
  \emph{Inf. Sci.}, vol. 311, pp. 18--38, Aug. 2015, doi:
  \url{10.1016/j.ins.2015.03.040}.

\bibitem{blitzer2007biographies}
J.~Blitzer, M.~Dredze, and F.~Pereira, ``Biographies, bollywood, boom-boxes and
  blenders: Domain adaptation for sentiment classification,'' in \emph{Proc.
  45th Annu. Meeting Association Computational Linguistics}, Prague, Czech
  Republic, Jun. 2007, pp. 440--447.

\bibitem{pan2010cross}
S.~J. Pan, X.~Ni, J.-T. Sun, Q.~Yang, and Z.~Chen, ``Cross-domain sentiment
  classification via spectral feature alignment,'' in \emph{Proc. 19th Int.
  Conf. World Wide Web (WWW '10)}, Wanchai, Hong Kong, Apr. 26, 2010, pp.
  751--760.

\bibitem{goldberg2006seeing}
A.~B. Goldberg and X.~Zhu, ``Seeing stars when there aren't many stars:
  Graph-based semi-supervised learning for sentiment categorization,'' in
  \emph{Proc. 1st Workshop Graph Based Methods for Natural Language
  Processing}, Stroudsburg, PA, USA, Jun. 2006, pp. 45--52, doi:
  \url{10.3115/1654758.1654769}.

\bibitem{mudinas2018bootstrap}
A.~Mudinas, D.~Zhang, and M.~Levene, ``Bootstrap domain-specific sentiment
  classifiers from unlabeled corpora,'' \emph{Trans. Assoc. Comput.
  Linguistics}, vol.~6, pp. 269--285, Jan. 2018, doi:
  \url{10.1162/tacl_a_00020}.

\bibitem{muhammad2016contextual}
A.~Muhammad, N.~Wiratunga, and R.~Lothian, ``Contextual sentiment analysis for
  social media genres,'' \emph{Knowl.-Based Syst.}, vol. 108, pp. 92--101, Sep.
  2016, doi: \url{10.1016/j.knosys.2016.05.032}.

\bibitem{taboada2011lexicon}
M.~Taboada, J.~Brooke, M.~Tofiloski, K.~Voll, and M.~Stede, ``Lexicon-based
  methods for sentiment analysis,'' \emph{Comp. Linguistics}, vol.~37, no.~2,
  pp. 267--307, Jun. 2011, doi: \url{10.1162/COLI_a_00049}.

\bibitem{ribeiro2016sentibench}
F.~N. Ribeiro, M.~Ara{\'u}jo, P.~Gon{\c{c}}alves, M.~A. Gon{\c{c}}alves, and
  F.~Benevenuto, ``Sentibench-a benchmark comparison of state-of-the-practice
  sentiment analysis methods,'' \emph{EPJ Data Sci.}, vol.~5, no.~1, p.~23,
  Dec. 2016, doi: \url{10.1140/epjds/s13688-016-0085-1}.

\bibitem{thelwall2012sentiment}
M.~Thelwall, K.~Buckley, and G.~Paltoglou, ``Sentiment strength detection for
  the social web,'' \emph{J. Amer. Soc. Inf. Sci. \& Technol.}, vol.~63, no.~1,
  pp. 163--173, Jan. 2012, doi: \url{10.1002/asi.21662}.

\bibitem{thelwall2010sentiment}
M.~Thelwall, K.~Buckley, G.~Paltoglou, D.~Cai, and A.~Kappas, ``Sentiment
  strength detection in short informal text,'' \emph{J. Amer. Soc. Inf. Sci. \&
  Technol.}, vol.~61, no.~12, pp. 2544--2558, Dec. 2010, doi:
  \url{10.1002/asi.21416}.

\bibitem{haselmayer2017sentiment}
M.~Haselmayer and M.~Jenny, ``Sentiment analysis of political communication:
  Combining a dictionary approach with crowdcoding,'' \emph{Qual. \& Quantity},
  vol.~51, no.~6, pp. 2623--2646, Nov. 2017, doi:
  \url{10.1007/s11135-016-0412-4}.

\bibitem{novielli2018benchmark}
N.~Novielli, D.~Girardi, and F.~Lanubile, ``A benchmark study on sentiment
  analysis for software engineering research,'' in \emph{IEEE/ACM 15th Int.
  Conf. on Mining Software Repositories (MSR '18)}, Gothenburg, Sweden, May 27,
  2018, pp. 364--375, doi: \url{10.1145/3196398.3196403}.

\bibitem{lin2018sentiment}
B.~Lin, F.~Zampetti, G.~Bavota, M.~Di~Penta, M.~Lanza, and R.~Oliveto,
  ``Sentiment analysis for software engineering: How far can we go?'' in
  \emph{IEEE/ACM 40th Int. Conf. Software Engineering (ICSE '18)}, Gothenburg,
  Sweden, May 27, 2018, pp. 94--104, doi: \url{10.1145/3180155.3180195}.

\bibitem{feldman2013techniques}
R.~Feldman, ``Techniques and applications for sentiment analysis,'' \emph{Comm.
  ACM}, vol.~56, no.~4, pp. 82--89, Apr. 2013, doi:
  \url{10.1145/2436256.2436274}.

\bibitem{cambria2013new}
E.~Cambria, B.~Schuller, Y.~Xia, and C.~Havasi, ``New avenues in opinion mining
  and sentiment analysis,'' \emph{IEEE Intell. Syst.}, vol.~28, no.~2, pp.
  15--21, Feb. 2013, doi: \url{10.1109/MIS.2013.30}.

\bibitem{duers2017learner}
L.~E. Duers, ``The learner as co-creator: A new peer review and self-assessment
  feedback form created by student nurses,'' \emph{Nurse Educ. Today}, vol.~58,
  pp. 47--52, 2017, doi: \url{10.1016/j.nedt.2017.08.002}.

\bibitem{wen2014sentiment}
M.~Wen, D.~Yang, and C.~Rose, ``Sentiment analysis in {MOOC} discussion forums:
  What does it tell us?'' in \emph{Educational Data Mining (EDM '14)}, London,
  U.K., Jul. 4, 2014.

\bibitem{chaplot2015predicting}
D.~S. Chaplot, E.~Rhim, and J.~Kim, ``Predicting student attrition in {MOOCs}
  using sentiment analysis and neural networks,'' in \emph{Artificial
  Intelligence in Education Workshop (AIED '15)}, vol.~53, Madrid, Spain, Jun.
  2015, pp. 54--57.

\bibitem{esparza2016proposal}
G.~G. Esparza, A.~P. D{\'\i}az, J.~Canul-Reich, C.~A. De-Luna, and J.~Ponce,
  ``Proposal of a sentiment analysis model in tweets for improvement of the
  teaching-learning process in the classroom using a corpus of subjectivity,''
  \emph{Int. J. Combinatorial Optim. Problems and Inform.}, vol.~7, no.~2, pp.
  22--34, May 2016.

\bibitem{pousada2017towards}
M.~Pousada, S.~Caball{\'e}, J.~Conesa, A.~Bertr{\'a}n,
  B.~G{\'o}mez-Z{\'u}{\~n}iga, E.~Hern{\'a}ndez, M.~Armayones, and J.~Mor{\'e},
  ``Towards a web-based teaching tool to measure and represent the emotional
  climate of virtual classrooms,'' in \emph{Int. Conf. Emerging
  Internetworking, Data \& Web Technologies}, Wuhan, China, Jun. 10, 2017, pp.
  314--327, doi: \url{10.1007/978-3-319-59463-7_32}.

\bibitem{schubert2018measuring}
M.~Schubert, D.~Durruty, and D.~A. Joyner, ``Measuring learner tone and
  sentiment at scale via text analysis of forum posts,'' in \emph{Proc. 8th
  Edition Int. Workshop Personalization Approaches Learning Environments
  (PALE)}, London, U.K., Jun. 27--30, 2018.

\bibitem{clarizia2018learning}
F.~Clarizia, F.~Colace, M.~De~Santo, M.~Lombardi, F.~Pascale, and
  A.~Pietrosanto, ``E-learning and sentiment analysis: A case study,'' in
  \emph{Proc. 6th Int. Conf. Information and Education Technology}, London,
  U.K., Jan. 6, 2018, pp. 111--118, doi: \url{10.1145/3178158.3178181}.

\bibitem{balahadia2016teacher}
F.~F. Balahadia, M.~C.~G. Fernando, and I.~C. Juanatas, ``Teacher's performance
  evaluation tool using opinion mining with sentiment analysis,'' in \emph{IEEE
  Region 10 Symposium (TENSYMP '16)}, Sanur, Indonesia, May 9, 2016, pp.
  95--98, doi: \url{10.1109/TENCONSpring.2016.7519384}.

\bibitem{rani2017sentiment}
S.~Rani and P.~Kumar, ``A sentiment analysis system to improve teaching and
  learning,'' \emph{Comput.}, vol.~50, no.~5, pp. 36--43, 2017, doi:
  \url{10.1109/MC.2017.133}.

\bibitem{lalata2019sentiment}
J.-a.~P. Lalata, B.~Gerardo, and R.~Medina, ``A sentiment analysis model for
  faculty comment evaluation using ensemble machine learning algorithms,'' in
  \emph{Proc. 2019 Int. Conf. Big Data Engineering}, Jun. 11, 2019, pp. 68--73,
  doi: \url{10.1145/3341620.3341638}.

\bibitem{fleming2018exploring}
M.~Fleming, B.~Coulter, and S.~Weaver, ``Exploring the student experience of
  industry placements using sentiment analysis,'' in \emph{29th Australasian
  Association for Engineering Education Conf. (AAEE '18)}.\hskip 1em plus 0.5em
  minus 0.4em\relax Hamilton, New Zealand: Engineers Australia, 2018, p. 213.

\bibitem{lynch2015soft}
T.~L. Lynch, ``Soft (a) ware in the english classroom: Spreadsheets and
  sinners: How and why english teachers can claim their rightful place in stem
  education,'' \emph{The English J.}, vol. 104, no.~5, pp. 98--101, May 2015.

\bibitem{gehringer2019board}
E.~F. Gehringer, ``Board 60: Peerlogic: Web services for peer assessment,'' in
  \emph{Proc. 2019 ASEE Annu. Conf. \& Expo.}, Tampa, FL, USA, Jun. 15, 2019.

\bibitem{gilbert2014vader}
\BIBentryALTinterwordspacing
C.~Gilbert and E.~Hutto, ``Vader: A parsimonious rule-based model for sentiment
  analysis of social media text,'' in \emph{8th Int. Conf. on Weblogs and
  Social Media (ICWSM-14)}, Oxford, U.K., May 16, 2014. [Online]. Available:
  \url{http://comp. social. gatech. edu/papers/icwsm14. vader. hutto. pdf}
\BIBentrySTDinterwordspacing

\bibitem{beasley2021domain}
Z.~J. Beasley and L.~A. Piegl, ``A domain-dependent lexicon to augment cad peer
  review,'' \emph{Comput.-Aided Des. and Appl.}, vol.~18, no.~1, pp. 186--198,
  May 2021, doi: \url{10.14733/cadaps.2021.186-198}.

\bibitem{piech2013tuned}
\BIBentryALTinterwordspacing
C.~Piech, J.~Huang, Z.~Chen, C.~Do, A.~Ng, and D.~Koller, ``Tuned models of
  peer assessment in {MOOCs},'' Jul. 2013. [Online]. Available:
  \url{arXiv:1307.2579}
\BIBentrySTDinterwordspacing

\bibitem{piegl2019assessing}
L.~A. Piegl, Z.~J. Beasley, and P.~Rosen, ``Assessing student design work using
  the intelligence of the crowd,'' in \emph{Proc. Computer-Aided Design Conf.
  and Exhibition (CAD '19)}, Changi, Singapore, Jun. 24--26, 2019, pp.
  117--121, doi: \url{10.14733/cadconfP.2019.117-121}.

\bibitem{belinkov2019analysis}
Y.~Belinkov and J.~Glass, ``Analysis methods in neural language processing: A
  survey,'' \emph{Trans. Assoc. Comput. Linguistics}, vol.~7, pp. 49--72, Aug.
  2019, doi: \url{10.1162/tacl_a_00254}.

\bibitem{yang2019xlnet}
\BIBentryALTinterwordspacing
Z.~Yang, Z.~Dai, Y.~Yang, J.~Carbonell, R.~Salakhutdinov, and Q.~V. Le,
  ``Xlnet: Generalized autoregressive pretraining for language understanding,''
  2019. [Online]. Available: \url{arXiv:1906.08237}
\BIBentrySTDinterwordspacing

\bibitem{afinn}
\BIBentryALTinterwordspacing
F.~{\AA}. Nielsen, ``A new anew: Evaluation of a word list for sentiment
  analysis in microblogs,'' Mar. 2011. [Online]. Available:
  \url{arXiv:1103.2903}
\BIBentrySTDinterwordspacing

\bibitem{pang2008opinion}
B.~Pang and L.~Lee, ``Opinion mining and sentiment analysis,'' \emph{Found. and
  Trends{\textregistered} Inf. Retrieval}, vol.~2, no. 1--2, pp. 1--135, Mar.
  2008, doi: \url{10.1561/1500000011}.

\bibitem{hatzivassiloglou1997predicting}
V.~Hatzivassiloglou and K.~R. McKeown, ``Predicting the semantic orientation of
  adjectives,'' in \emph{Proc. 35th Annu. Meeting Association Computational
  Linguistics and 8th Conf. European Chapter Association Computational
  Linguistics}, Madrid, Spain, Jul. 1997, pp. 174--181, doi:
  \url{10.3115/976909.979640}.

\bibitem{turney2002thumbs}
P.~D. Turney, ``Thumbs up or thumbs down?: Semantic orientation applied to
  unsupervised classification of reviews,'' in \emph{Proc. 40th Annu. Meeting
  Association Computational Linguistics}, Philadelphia, PA, USA, Dec. 11, 2002,
  pp. 417--424, doi: \url{10.3115/1118693.1118704}.

\bibitem{hu2004mining}
M.~Hu and B.~Liu, ``Mining and summarizing customer reviews,'' in \emph{Proc.
  10th ACM Int. Conf. Knowledge Discovery and Data Mining (SIGKDD '04)},
  Seattle, WA, USA, Aug. 22, 2004, pp. 168--177, doi:
  \url{10.1145/1014052.1014073}.

\bibitem{beasley2020crowd}
\BIBentryALTinterwordspacing
Z.~Beasley, ``Crowd-sourced plagiarism detection of essays,'' in \emph{Proc.
  Southeast Int. Center for Academic Integrity Conf. (ICAI '20)}, Tampa, FL,
  USA, Jun. 6--11, 2020. [Online]. Available:
  \url{https://scholarcommons.usf.edu/cgi/viewcontent.cgi?article=1002&context=southeast-icai-conference}
\BIBentrySTDinterwordspacing

\bibitem{beasley2020helps}
Z.~J. Beasley and L.~A. Piegl, ``Helps: A domain-specific lexicon for cad peer
  review,'' in \emph{Proc. CAD Conf. and Exhibition}, Barcelona, Spain, Jul.
  6--8, 2020, pp. 21--25, doi: \url{10.14733/cadconfP.2020.21-25}.

\bibitem{blair2008building}
S.~Blair-Goldensohn, K.~Hannan, R.~McDonald, T.~Neylon, G.~A. Reis, and
  J.~Reynar, ``Building a sentiment summarizer for local service reviews,'' in
  \emph{Proc. World Wide Web Workshop NLP Challenges Information Explosion Era
  (NLPIX '08)}, vol.~14, Beijing, China, Apr. 22, 2008, pp. 339--348.

\bibitem{beasley2020sentiment}
\BIBentryALTinterwordspacing
Z.~J. Beasley, ``Sentiment analysis in peer review,'' Ph.D. dissertation,
  University of South Florida, 2020. [Online]. Available:
  \url{https://scholarcommons.usf.edu/etd/8160}
\BIBentrySTDinterwordspacing

\bibitem{potts2010negativity}
C.~Potts, ``On the negativity of negation,'' \emph{Semantics and Linguistic
  Theor.}, vol.~20, pp. 636--659, Aug. 2010, doi:
  \url{10.3765/salt.v20i0.2565}.

\bibitem{wiegand2010survey}
M.~Wiegand, A.~Balahur, B.~Roth, D.~Klakow, and A.~Montoyo, ``A survey on the
  role of negation in sentiment analysis,'' in \emph{Proc. Workshop Negation
  and Speculation in Natural Language Processing}, Uppsala, Sweden, Jul. 2010,
  pp. 60--68.

\bibitem{pang2002thumbs}
B.~Pang, L.~Lee, and S.~Vaithyanathan, ``Thumbs up?: sentiment classification
  using machine learning techniques,'' in \emph{Proc. ACL Conf. Empirical
  Methods in Natural Language Processing}, Philadelphia, PA, USA, May 28, 2002,
  pp. 79--86.

\bibitem{neviarouskaya2011sentiful}
A.~Neviarouskaya, H.~Prendinger, and M.~Ishizuka, ``Sentiful: A lexicon for
  sentiment analysis,'' \emph{IEEE Trans. Affect. Comput.}, vol.~2, no.~1, pp.
  22--36, Feb. 2011, doi: \url{10.1109/T-AFFC.2011.1}.

\bibitem{wilson2005recognizing}
T.~Wilson, J.~Wiebe, and P.~Hoffmann, ``Recognizing contextual polarity in
  phrase-level sentiment analysis,'' in \emph{Proc. Human Language Technology
  Conf. and Conf. Empirical Methods in Natural Language Processing}, Vancouver,
  Canada, Oct. 2005, doi: \url{10.3115/1220575.1220619}.

\bibitem{dave2003mining}
K.~Dave, S.~Lawrence, and D.~M. Pennock, ``Mining the peanut gallery: Opinion
  extraction and semantic classification of product reviews,'' in \emph{Proc.
  12th Int. Conf. World Wide Web (WWW '03)}, Budapest, Hungary, May 20, 2003,
  pp. 519--528, doi: \url{10.1145/775152.775226}.

\bibitem{fang2015sentiment}
X.~Fang and J.~Zhan, ``Sentiment analysis using product review data,'' \emph{J.
  Big Data}, vol.~2, no.~1, p.~5, Dec. 2015, doi:
  \url{10.1186/s40537-015-0015-2}.

\bibitem{bradley1999affective}
M.~M. Bradley and P.~J. Lang, ``Affective norms for english words (anew):
  Instruction manual and affective ratings,'' Center Research Psychophysiology,
  Univ. Florida, Gainsville, FL, USA, Tech. Rep. C-1, Jan. 1999.

\bibitem{wu2016slangsd}
\BIBentryALTinterwordspacing
L.~Wu, F.~Morstatter, and H.~Liu, ``Slangsd: Building and using a sentiment
  dictionary of slang words for short-text sentiment classification,'' Aug.
  2016. [Online]. Available: \url{arXiv:1608.05129}
\BIBentrySTDinterwordspacing

\bibitem{deng2015mpqa}
L.~Deng and J.~Wiebe, ``Mpqa 3.0: An entity/event-level sentiment corpus,'' in
  \emph{Proc. 2015 Conf. North American Chapter Association for Computational
  Linguistics: Human Language Technologies}, Denver, CO, USA, 2015, pp.
  1323--1328, doi: \url{10.3115/v1/N15-1146}.

\bibitem{baccianella2010sentiwordnet}
S.~Baccianella, A.~Esuli, and F.~Sebastiani, ``Sentiwordnet 3.0: An enhanced
  lexical resource for sentiment analysis and opinion mining,'' \emph{LREC},
  vol.~10, no. 2010, pp. 2200--2204, May 2010.

\bibitem{hoffmann2019benchmarking}
F.~Hoffmann, T.~Bertram, R.~Mikut, M.~Reischl, and O.~Nelles, ``Benchmarking in
  classification and regression,'' \emph{Wiley Interdisciplinary Rev.: Data
  Mining and Knowl. Discovery}, vol.~9, no.~5, p. e1318, Sep. 2019, doi:
  \url{10.1002/widm.1318}.

\bibitem{beasley2020leveraging}
Z.~Beasley, A.~Friedman, L.~Pieg, and P.~Rosen, ``Leveraging peer feedback to
  improve visualization education,'' in \emph{IEEE Pacific Visualization
  Symposium (PacificVis '20)}, Jun. 3, 2020, pp. 146--155, doi:
  \url{10.1109/PacificVis48177.2020.1261}.

\end{thebibliography}

\end{document}